\documentclass[twocolumn,prb,amsmath,amssymb,floatfix,superscriptaddress,showpacs]{revtex4}
\usepackage{color}
\usepackage{graphicx}
\usepackage{dcolumn}
\usepackage{bm}
\usepackage{times}
\begin{document}


\title{Freezing of the local dynamics in the relaxor ferroelectric PZN-4.5\%PT}

\author{Zhijun Xu}
\affiliation{Condensed Matter Physics and Materials Science
Department, Brookhaven National Laboratory, Upton, New York 11973,
USA} \affiliation{Department of Physics, City College of New York,
New York, New York 10033, USA}
\author{Jinsheng Wen}
\affiliation{Condensed Matter Physics and Materials Science Department, Brookhaven National Laboratory, Upton, New York 11973,
USA}\affiliation{Department of Materials Science and Engineering, Stony Brook University, Stony Brook, New York 11794, USA}
\author{E. Mamontov}
\affiliation{Neutron Scattering Science Division, Oak Ridge National
Laboratory, Oak Ridge, Tennessee 37831, USA}
\author{C. Stock}
\affiliation{NIST Center for Neutron Research, National Institute of Standards and Technology, Gaithersburg, Maryland 20899, and
Indiana University Cyclotron Facility, Bloomington, Indiana 47404, USA}
\author{P. M. Gehring}
\affiliation{NIST Center for Neutron Research, National Institute of Standards and Technology, Gaithersburg, Maryland 20899, USA}
\author{Guangyong Xu}
\affiliation{Condensed Matter Physics and Materials Science
Department, Brookhaven National Laboratory, Upton, New York 11973,
USA}

\date{\today}

\begin{abstract}
We report measurements of the neutron diffuse scattering in a single
crystal of the relaxor ferroelectric material
95.5\%Pb(Zn$_{1/3}$Nb$_{2/3}$)O$_3$-4.5\%PbTiO$_3$ (PZN-4.5\%PT). We
show that the diffuse scattering at high temperatures has a
quasielastic component with energy width $\agt 0.1$\,meV.  On
cooling the total diffuse scattering intensity increases, but the
intensity and the energy width of the quasielastic component
gradually diminish.  At 50\,K the diffuse scattering is completely
static (i.\ e.\ the energy width lies within the limits of our
instrumental resolution).  This suggests that the dynamics of the
short-range correlated atomic displacements associated with the
diffuse scattering freeze at low temperature.  We find that this
depends on the wave vector $q$ as the quasielastic diffuse
scattering intensities associated with $\langle001\rangle$ (T1-type)
and $\langle110\rangle$ (T2-type) atomic displacements vary
differently with temperature and electric field.
\end{abstract}


\maketitle

\section{Introduction}

The physics of lead-based, perovskite, relaxor systems (PbBO$_3$),
such as Pb(Zn$_{1/3}$Nb$_{2/3}$)O$_3$ (PZN), and
Pb(Mg$_{1/3}$Nb$_{2/3}$)O$_3$ (PMN), are complicated by the random
fields generated by the charge disorder on the B-site.
Nanometer-scale polar clusters, or ``polar nano-regions'' (PNRs),
are widely believed to form at the Burns temperature
$T_{d}$,~\cite{burns1983} which is typically a few hundred degrees
above the Curie temperature $T_C$. Many unusual bulk properties of
relaxor systems have been attributed to these
PNRs~\cite{cross1987,Burton2006,Gehring2009,Matsuura2006,gxu2008nm},
and consequently they have been the focus of extensive study.
Various models of the short-range correlated (i.\ e.\ local) atomic
displacements that define the PNRs have been proposed based on
extensive neutron and x-ray scattering studies of these and other
related relaxor
systems.~\cite{gxu2006nm,gxu20043d,Welberry2005,Welberry2006,Gvasaliya2009,Jeong2005,Vakhrushev1995,hirota2002}
Our most recent work~\cite{zxu2010relaxor1} shows that the diffuse
scattering measured in a specific region of reciprocal space varies
strongly in the presence of an electric field {\bf E} oriented along
[111], but not for {\bf E} oriented along [100], while the opposite
behavior is observed for the diffuse scattering measured in a nearby
region of reciprocal space.  This suggests that the local atomic
displacements comprising the PNRs may be composed of two distinct
components that give rise to two distinct, but overlapping, diffuse
scattering distributions that are located near every Brillouin zone
center.  In our model the local atomic displacements along $\langle110\rangle$
are responsible for the well-known butterfly-shaped diffuse
scattering,~\cite{You1997,Hlinka2003jpcm,Hiraka2004prb,gxu2004prb,gxu2006nm,Stock2007}
which responds strongly to {\bf E} along [111]; we then speculate
that local atomic displacements along $\langle001\rangle$ could
produce a differently shaped diffuse scattering distribution that
would instead respond strongly to {\bf E} along [100].  Following
previously defined nomenclature, we shall refer to the first as
T2-diffuse scattering because of its similarity to T2 transverse
acoustic (TA) phonon modes, which are polarized along $\langle
110\rangle$,~\cite{gxu2008nm} and we shall refer to the second as
T1-diffuse scattering by analogy to T1 phonon modes, which are
polarized along $\langle001\rangle$.~\cite{Gehring2004prb}

Scattering methods are essential tools for mapping out the wave
vector ({\bf Q}) dependence of the diffuse scattering in relaxors,
which in turn provides key structural information about the atomic
displacements associated with the PNRs as well as the length scales
over which these displacements are correlated.  One can also obtain
information about the energy/time scales associated with the PNRs.
Neutron-based methods can easily distinguish between static and
thermal diffuse scattering.  However, depending on the temperature,
the diffuse scattering from the PNRs can exhibit both elastic and
quasielastic components; this has been conclusively demonstrated by
neutron measurements made with very high energy
resolution.~\cite{Gvasaliya2004,Hiraka2004prb,Gehring2009,Stock2010,zxu2010relaxor1}
Recent neutron spin echo measurements have shown that the PNRs in
both PMN and the related relaxor system PZN-4.5\%PT (a solid
solution of PZN and PbTiO$_3$) display relaxational dynamics at high
temperatures with a typical lifetime of around $0.0042$\,ns, which
corresponds to an energy width of $0.16$\,meV.  These dynamic, local
structures gradually freeze on cooling and become entirely static at
sufficiently low temperatures.

In this paper, we discuss detailed measurements of the T1- and T2-diffuse
scattering components made simultaneously on the same single crystal of
PZN-4.5\%PT.  We find that the dynamics of both components follow
the same trend - freezing with cooling.  One difference is that the
T1-component exhibits a narrower energy width, i.\ e.\ a longer life time,
than does the T2-component measured at the same temperature.  An external
field applied along [001] does not affect the T2-component, but it
significantly reduces the intensity of the T1-component near T$_C$.  These
results confirm that subtle differences exist between the dynamics of PNRs
measured at two different wave vectors within the same Brillouin zone and
lend support to the concept that the PNRs are composed of two distinct
components.  Possible connections between the local dynamics and bulk lattice
dynamics (phonons) are discussed.

\begin{figure}[ht]
\includegraphics[width=\linewidth]{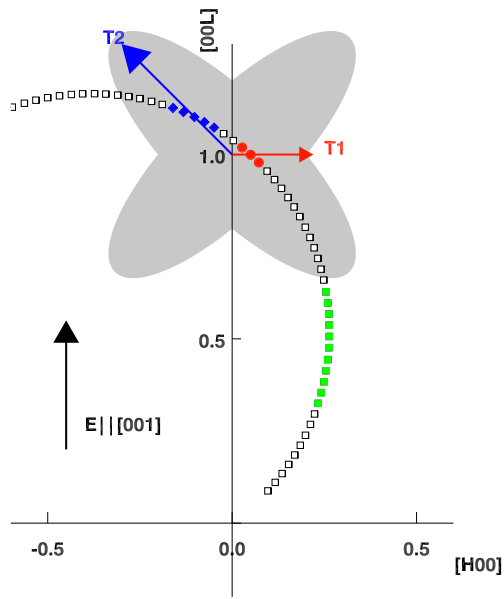}
\caption{(Color online) Schematic diagram of the scattering geometry
in which measurements were made in the (H0L) plane with an electric
field applied along [001]. The grey butterfly-shaped region represents
constant-intensity contours of the T2-diffuse scattering centered at
{\bf Q}=(001). The small squares show the locations in {\bf Q} of the
BASIS detectors for elastically ($\hbar\omega=0$) scattered neutrons.}
\label{fig:1}
\end{figure}

\begin{figure}[ht]
\includegraphics[width=\linewidth]{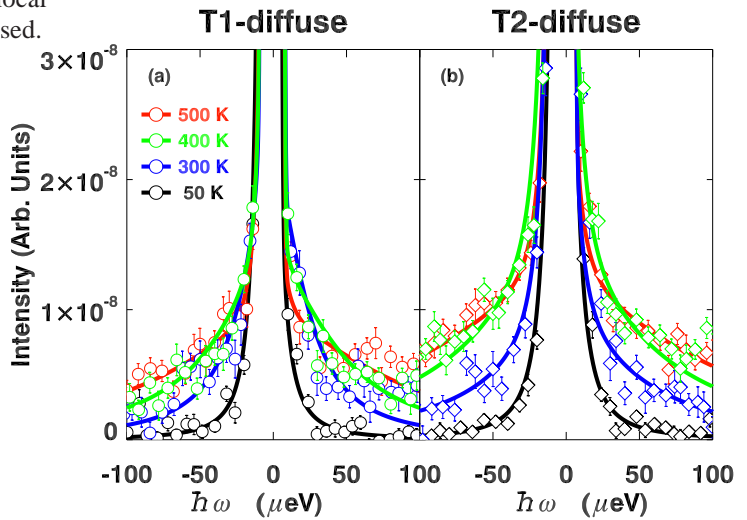}
\caption{(Color online) Diffuse scattering intensities are plotted versus
$\hbar\omega$ at 500\,K (red), 400\,K (green), 300\,K (blue), and 50\,K
(black) for the (a) T1-component, measured at {\bf Q} = (0.05,0,1), and the
(b) T2-component, measured at {\bf Q} = (-0.1,0,1.1).  The solid lines are
based on the least square fits to the data described in the text.  The error
bars represent the square root of the number of counts.}
\label{fig:2}
\end{figure}

\begin{figure}[ht]
\includegraphics[width=\linewidth]{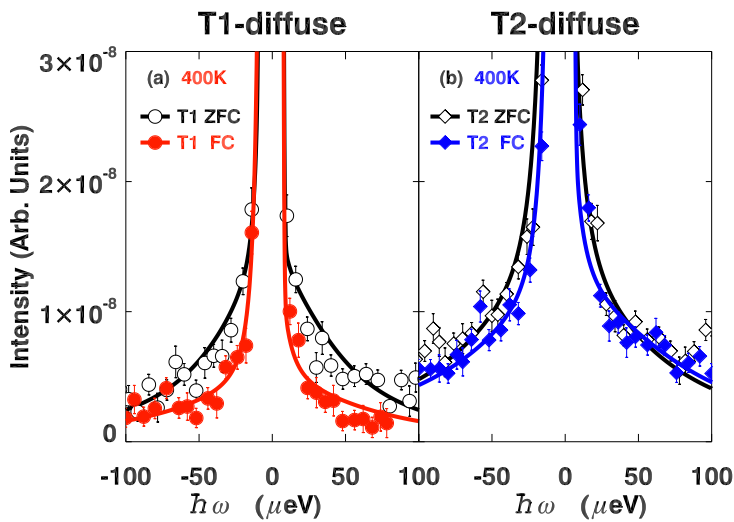}
\caption{(Color online) Diffuse scattering intensities are plotted versus
$\hbar\omega$ at 400\,K after zero-field cooling (ZFC: black, open symbols)
and field cooling (FC:  red, solid symbols).  Data are shown for the (a)
T1-component, measured at {\bf Q} = (0.05,0,1), and the (b) T2-component,
measured at {\bf Q} = (-0.1,0,1.1). The solid lines are based on the least
square fits to the data described in the text.  The error bars represent
the square root of the number of counts.}
\label{fig:3}
\end{figure}

\begin{figure}[ht]
\includegraphics[width=\linewidth]{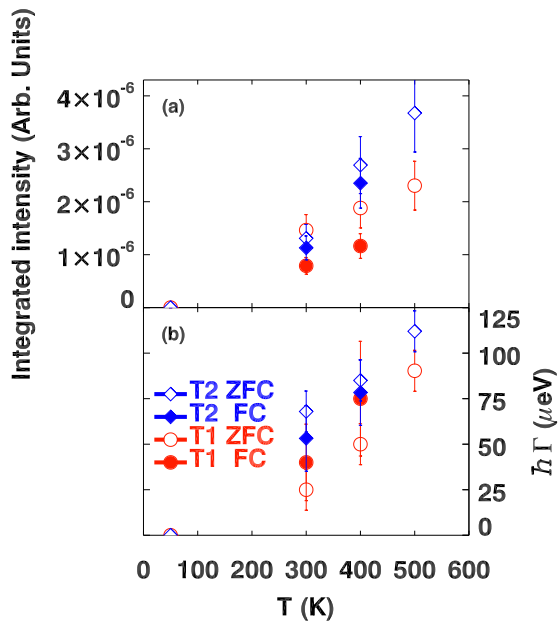}
\caption{(Color online) (a) Energy width $\hbar\Gamma$ (half width at half
maximum) of the quasielastic diffuse scattering versus temperature for the
T1-component (ZFC = red, open circles; FC = red, solid circles) and the
T2-component (ZFC = blue, open diamonds; FC = blue, solid diamonds).
(b) Temperature dependence of the integrated intensities of the quasielastic
diffuse scattering.}
\label{fig:4}
\end{figure}

\section{Experimental Details}

The sample studied in this experiment is a rectangular single crystal
of PZN-4.5\%PT with \{100\} surfaces and dimensions of
10$\times$10$\times$3\,mm$^3$.  The sample has a cubic lattice
spacing of $a = 4.05$\,\AA\ at 300\,K; thus 1\,r.l.u. (reciprocal
lattice unit) equals $2\pi/a = 1.55$\,\AA$^{-1}$.  Cr/Au electrodes
were sputtered onto the two largest opposing crystal surfaces.  The
Curie temperature of this compound $T_C \sim 475$\,K.  This is
manifested by a release of extinction that results in a more than
two-fold increase in the intensity of the (100) Bragg peak on
cooling.~\cite{zxu2010relaxor1}

Neutron diffuse scattering measurements were performed on the BASIS
backscattering spectrometer, which is located at the Oak Ridge National
Laboratory Spallation Neutron Source (SNS).  A large bank of Si(111)
crystals is employed as analyzer to reflect those neutrons scattered by
the sample that have a final energy of Ef=2.082\,meV.  The incident
neutron energy bandwidth is centered around the same energy using a
series of bandwidth choppers. The instrumental energy resolution is
about 1.5\,$\mu$eV half-width at half-maximum (HWHM) for elastically
($\hbar\omega=0$) scattered neutrons. The $a$-axis of the crystal was
aligned so that it formed a 54\,$^{\circ}$ angle with respect to the
incident beam.  In this configuration the BASIS detectors collected
quasielastic scattering (i.\ e.\ centered around $\hbar\omega=0$)
intensities at the reciprocal space locations shown in Fig.~\ref{fig:1}.
Given the large detector coverage provided by BASIS we were able to
measure the diffuse scattering intensities at many different ${\bf Q}$
values at the same time.  Our measurements show that this spectrometer
is well-suited to the study of quasielastic scattering in the $\mu$eV
energy range from single crystal samples.  An external electric field
$E=1$\,kV/cm was applied along [001] above 550\,K during all of the
field-cooled (FC) measurements.

\section{Results and Discussion}

We performed a series of measurements to characterize the dynamics of the
diffuse scattering in the (H0L) plane under different temperatures and
field-cooling conditions. Based on the area detector design of BASIS,
the group of detectors marked in blue in Fig.~\ref{fig:1} were summed
to give the intensity in a small neighborhood of ${\bf Q}=(-0.10,0,1.10)$,
which corresponds to the T2-component.  Similarly the group of detectors
shown in red should reflect intensities near ${\bf Q} =(0.05,0,1)$, which
corresponds to the T1-component.  The group of detectors shown in green
are located far enough from the Bragg peak that they may be summed and
used to measure the background; the intensities collected by this group
of detectors exhibited no measurable {\bf Q} or temperature dependence.

The T1 diffuse scattering intensities measured by the red detectors
are plotted in panel (a) of Fig.~\ref{fig:2} after first subtracting
out the background intensities collected by the green detectors.  At
50~K the diffuse scattering lineshape is resolution-limited; hence
all quasielastic processes occurring on instrumentally-accessible
time scales are frozen. These data are well-described by the sum of
a Gaussian and Lorentzian function of energy, and they represent the
total (static) scattering response of the system, which is composed
of the static diffuse scattering plus any incoherent scattering.
This curve is slightly asymmetric as explained in
Ref.~\onlinecite{Mamontov2011}, and was then used to model the
instrumental energy resolution function at all temperatures. Data
measured above 50\,K were fit to the same Gaussian and Lorentzian
functions times an overall scale factor, and then added to another
Lorentzian function that was used to parameterize the quasielastic
component. The T2-diffuse scattering intensities are plotted in
panel (b) of Fig.~\ref{fig:2}; the fittings for these data were
performed in the same manner as done for the T1-component data.

On cooling quasielastic scattering from both the T1- and
T2-components appears at temperatures well above $T_{C}$.  At 500\,K
both components display slow dynamics characterized by HWHM energy
widths of $\hbar\Gamma(T1) \sim 0.09(\pm 0.005)$\,meV and
$\hbar\Gamma(T2) 0.11(\pm 0.005)$\,meV. These values are in
reasonably good agreement with those determined by our recent
spin-echo measurements.~\cite{Stock2010,zxu2010relaxor1}  At lower
temperatures both energy widths narrow, as shown in panel (b) of
Fig.~\ref{fig:4}.  The quasielastic scattering intensity also
decreases monotonically from 500\,K to 300\,K.  Near 50\,K the
diffuse scattering becomes entirely static.  We note that for all
temperatures studied, the energy widths $\Gamma(T1)< \Gamma(T2)$,
which suggests that subtle differences exist between the local
dynamics associated with $\langle001\rangle$ and $\langle110\rangle$
oriented atomic displacements.  At a given temperature the
T2-component exhibits a shorter lifetime, but at sufficiently low
temperatures both are entirely static.

The integrated intensity of the quasielastic scattering associated
with both the T1- and T2-components (see panel (a) of
Fig.~\ref{fig:4}) also decreases on cooling, whereas the total
diffuse scattering intensity increases.  This confirms that the PNRs
become increasingly longer-lived as the temperature is lowered.  The
quasielastic scattering intensity of the T2-component is clearly
stronger than that of the T1-component, even though the T2-component
intensities are being measured at an average wave vector located
further from the Bragg peak.  We thus conclude that the T2-component
is the dominant contribution to the diffuse scattering intensities.

We have also examined the effects of an external electric field on
the quasielastic components of the T1- and T2-diffuse scattering. As
can be seen from panel (b) of Fig.~\ref{fig:3}, essentially no
change in the T2-component is observed when the system is cooled
from 550\,K to 400\,K in a 1\,kV/cm electric field applied along
[001].  This result is consistent with our previous
observations.~\cite{Wen2008apl,zxu2010relaxor1}  By contrast, the
data in panel (a) of Fig.~\ref{fig:3} demonstrate that the same
field greatly affects (reduces) the T1-component.  In order to
characterize these observations in greater detail, the integrated
intensities and the energy widths (HWHM) of the two quasielastic
components were extracted from fits to the total diffuse scattering
as previously described.  These quantities are plotted as a function
of temperature in Fig.~\ref{fig:4} for both ZFC and FC conditions.
We emphasize that the [001] field direction is parallel to the
direction of the local atomic displacements that are associated with
the T1-diffuse scattering measured near {\bf Q}$=(001)$.  As
expected, these data show that the quasielastic component of the
T1-diffuse scattering intensity is significantly weakened by the
[001] electric field, but that the same is not true for the
T2-component.~\cite{zxu2010relaxor1} The energy widths associated
with each component, on the other hand, have such large uncertainties
that no conclusion can be drawn about the respective electric field
dependence.  For example, at 400\,K the field-cooled
quasielastic energy width of the T1-component appears to be enhanced
relative to the zero-field cooled case, but the difference between the
two lies within our experimental error.  On cooling to 300\,K the
field effects appear to become less pronounced, but this trend also
lies within our experimental uncertainties.

Our finding that the dynamics associated with the T1-component are slower
than those associated with the T2-component may be understood when
interactions between the acoustic phonons and the PNRs
are taken into consideration.  Normally T2 phonons are softer than T1
phonons in PZN-$x$\%PT and PMN-$x$\%PT relaxors that exhibit
a rhombohedral ground state.~\cite{Swainson2009}  Previous work has
conclusively shown that whereas the relaxational T2-diffuse scattering
couples strongly to TA2 phonons,~\cite{gxu2008nm,Stock2005,Stock2012} the
coupling between the relaxational T1-diffuse scattering and the TA1
phonons is weaker~\cite{Stock2012}.
There are, however, also indications of a coupling between the T1-diffuse
and TO1 optic phonons.~\cite{zxu2010relaxor2} In other words, the
local atomic displacements within the PNRs can be affected by
phonons provided that they share the same polarization; these then
induce slow, local modes within the PNRs, which give rise to the quasielastic
scattering that are discussed in this paper. Along $\langle 110\rangle$
where the bulk phonons are softer and the coupling is stronger, the
local atomic displacements become more dynamic and therefore the
quasielastic component of the T2-diffuse scattering could have a larger
energy width. Our finding thus adds an important, new piece of
information to the already puzzling picture of competing and
coexisting local and long-range polar order in
relaxor systems~\cite{gxu2006prb} and deserves further study.

\section{Summary}

We have characterized the temperature and field dependence of the
quasielastic diffuse scattering measured from a single crystal sample of
the relaxor ferroelectric PZN-4.5\%PT using neutron backscattering, which
provides excellent energy resolution.  Our data show that both the T1- and
T2-components of the diffuse scattering exhibit quasielastic character at
high temperatures that diminishes gradually on cooling.  We also observe
differences between the dynamics of these two components, which provides
further evidence that, in addition to the well-known and more extensively
studied T2-component, a distinct, weaker, and lesser known T1-component is
also present, which is associated with local atomic displacements along
$\langle 001\rangle$.  Given that the T2-component has already been shown
to affect the polar properties of this relaxor system, it is quite
possible that the T1-component does so as well.  It thus merits study in
much greater detail in the future.

\section{Acknowledgments}

Financial support from the US Department of Energy under contract
No.\ DE-AC02-98CH10886 is gratefully acknowledged. This research at
the Oak Ridge National Laboratory Spallation Neutron Source was
sponsored by the Scientific User Facilities Division, Office of
Basic Energy Sciences, U.\ S.\ Department of Energy.


\end{document}